# Vacuum induced phonon transfer between two solid dielectric materials: Illustrating the case of Casimir force coupling

Younès Ezzahri* and Karl Joulain[§]

*Institut Pprime, Université de Poitiers-CNRS-ENSMA*

*2, Rue Pierre Brousse Bâtiment B25, TSA 41105*

*86073 Poitiers Cedex 9 France*

*[*]: younes.ezzahri@univ-poitiers.fr*

*[§]: karl.joulain@univ-poitiers.fr*

## ABSTRACT

The natural transition from the radiative regime to the conductive regime of heat transfer between two identical isotropic dielectric solid materials, is questioned by investigating the possibility of induced phonon transfer in vacuum. We describe the process in a general way assuming a certain phonon coupling mechanism between the two identical solids, then we particularly illustrate the case of coupling through Casimir force. We analyze how this mechanism of heat transfer compares and competes with the near field thermal radiation using a local model of the dielectric function. We show that the former mechanism can be very effective and even overpass the latter mechanism depending on the nature of the solid dielectric materials, the distance gap between them as well as the operating temperature regime.

# I. INTRODUCTION

Understanding and controlling heat transfer at very short length scales, has become very crucial and challenging in the last decade due to the continuous development in nanotechnology and the rapid evolution in the synthesis and fabrication of different materials at a nanometer scale.[1,2] At these scales, two heat transfer mechanisms become dominant, namely near field thermal radiation (mediated by photons) and interface conduction (mediated by phonons) between two solid materials.

The study of radiation heat transfer between two solid materials has become a topic of great interest in the last decade.[3,4] In addition to the purely fundamental aspect of the phenomenon, the interest was mainly motivated due to the increasing application potential in different technological domains, particularly in renewable energy sources such as in photovoltaics and thermophotovoltaics.[5,6]

In classical radiation theory, the radiation heat transfer between two solids is maximal when both solids behave as black bodies.[7] The situation changes radically when the separation distance between the two solids becomes comparable to or smaller than the dominant wavelength of the thermal radiation called Wien length ($\lambda_T = \hbar c / K_B T$), where $T$ is the absolute temperature, $c$ is the speed of light in vacuum, $\hbar$ is the reduced Planck constant and $K_B$ is Boltzmann constant. Other very interesting physical effects emerge such as heat transfer through tunneling of the electromagnetic evanescent waves. Due to the inclusion of these effects, radiation heat transfer between two solid materials increases enormously and becomes even orders of magnitude higher than the black body limit.[3,4] In addition, as the separation distance gets shorter and shorter for the two materials to mutually touch, a conductive heat transfer starts to take place through the new interface. Thus, a natural transition from the radiative regime to the conductive regime of heat transfer will occur as the separation distance between the two solid materials tends to zero. This transition is therefore intimately linked to the notion of the interface thermal resistance depending on the nature of the two solids materials (metal or dielectric).[8]

The must-occurrence of such a transition regime in heat transfer between two solid materials has raised a very fundamental question regarding the possibility of phonon tunneling through the separation gap between the two solids when the latter become very close to each other. When the gap is filled of nothing (vacuum), speaking of phonon tunneling can be very misleading. In fact, this terminology will be more respected if one considers phonons (acoustic or optic waves) that makes the surface vibrates. In fact, these waves could tunnel if the amplitude of their displacement is on the same order or higher than the gap distance. On the other hand, a bulk acoustic or optic phonon (elementary vibration inside a matter) cannot propagate in vacuum. Therefore, it will be more meaningful to speak of *induced phonon transfer*; elementary vibrations in one solid material will induce elementary vibrations in the other material and vice versa when the two are brought very close to each other so that a certain phonon coupling mechanism is established. There has been few works tackling this question imagining different coupling mechanisms for the occurrence of this phonon induction phenomenon by differentiating particularly the case of piezoelectric and nonpiezoelectric crystals.[9-13]

Indeed, the purpose of our present work is to investigate the possibility of such induced phonon transfer by considering a general phonon coupling mechanism between two solid materials. Then, we illustrate our approach assuming coupling through Casimir force.[14,15] This will be a generalization of the approach recently presented by Budaev and Bogy.[12,13] As a matter of fact, Casimir force is the most famous mechanical effect of vacuum fluctuations. The investigation of which has seen a rapidly growing activity in the last two decades due to its potential influence on the working of nanosystems and nanoscale structures such as micromachining devices and microelectromechanical systems (MEMS).[16] Casimir force depends on the gap distance between the two solids as well as on their optical properties, hence a local change of this distance due to any displacement of one side of the gap in the acoustic wave causes an excess pressure on the opposite side of the gap.[9]

## II. THEORY

A sketch of the situation under study is illustrated in Fig. 1. For simplicity sake, and without loss of generality, we will consider two identical nonmagnetic isotropic semi-infinite parallel plane solid materials both in a thermal equilibrium state at different temperatures, to be put in vacuum and separated by a gap distance $d$. The situation corresponds to a point junction case, through which the transport of phonons may be regarded as ballistic.[17] Each solid material is characterized by: (i) an atomic mass $m$, (ii) a spring constant $k$ corresponding to a harmonic potential and (iii) a dielectric permittivity function $\varepsilon$. The two solids are then connected via a certain phonon coupling mechanism described by a harmonic potential and represented by a spring coupling constant $k_{Coupling}$.

The calculation of the phonon heat flux density through the interface can be carried out using either the Scattering Boundary Method (SBM)[17] or NonEquilibrium Green's Function Method (NEGFM).[18,19] Within the harmonic approximation, the two methods have been shown to be equivalent and give the same results.[17] Besides, the SBM has the advantage of simplicity and can provide very closed-form analytical expressions for the phonon transmission function.[17] Thus, we choose herein to use the SBM to work out our analysis.

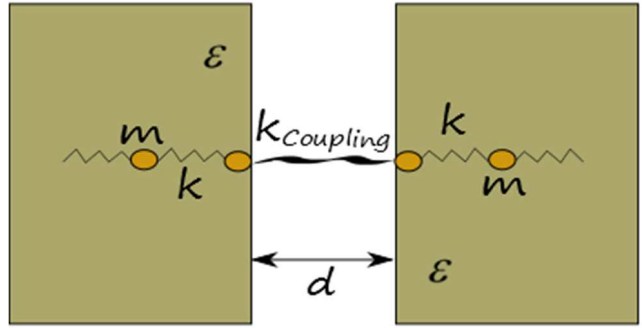

**Figure 1 :** Schematic illustration of the studied structure.

Using a Landauer formalism, one can show that the phononic thermal conductance through the interface takes the expression:[18,19]

$$\sigma_{Ph}(T,\omega_C,k_{Coupling}) = \frac{1}{2\pi} \int_0^{\omega_C} \tau_{3D}(\omega^2,\omega_C^2,k,k_{Coupling}) C_{Ph}(\omega,T) d\omega \quad (1)$$

where $\omega_C = 2\sqrt{k/m}$ is the cutoff frequency in the phonon dispersion relation inside each material, $C_{Ph}(\omega,T) = \hbar\omega[\partial n_0(\omega,T)/\partial T]$ represents the specific heat per normal phonon mode and $n_0(\omega,T) = [\exp(\hbar\omega/K_B T) - 1]^{-1}$ is the Planck equilibrium phonon distribution function.

The key step for the calculation of $\sigma_{Ph}(T)$ is the determination of the frequency dependent transmission function for the 3D configuration we are considering $\tau_{3D}(\omega^2,\omega_C^2,k,k_{Coupling})$. The latter gathers all the information about the nature of the phonon transport mechanisms. According to the SBM, the 1D configuration transmission function in the case of two identical solid harmonic chains can be written as:[17]

$$\tau_{1D}(\omega^2,\omega_C^2,k,k_{Coupling}) = \frac{k_{Coupling}^2(\omega_C^2 - \omega^2)}{k(k - 2k_{Coupling})\omega^2 + k_{Coupling}^2 \omega_C^2} \quad (2)$$

In order to obtain the transmission function in the 3D configuration, it is important to take into consideration the isotropy of the solid media and the conservation of the parallel momentum (wave-vector) relative to the phonon dispersion.[19] Therefore, within the frame work of the linear acoustic Debye theory, to go from 1D configuration to 3D configuration, one uses the relation:[19]

$$\tau_{3D}\left(\omega^2,\omega_C^2,k,k_{Coupling}\right) = \frac{1}{2\pi}\int_0^{\omega/v} \tau_{1D}\left(\omega^2-v^2q^2,\omega_C^2,k,k_{Coupling}\right)qdq \quad (3)$$

where $q$ is the parallel wave-vector and $v$ represents an average sound velocity that takes into account both longitudinal and transverse acoustic phonon polarizations $3/v^2 = 1/v_L^2 + 2/v_T^2$.

Eq. (3) can be worked out analytically, hence the transmission function in the 3D configuration of two identical solid materials is given by:

$$\tau_{3D}\left(\omega^2,\omega_C^2,k,k_{Coupling}\right) = \frac{k_{Coupling}^2\omega_C^2}{4\pi k^2\left(k-2k_{Coupling}\right)^2 v^2}\left\{\begin{array}{l}\left(k-k_{Coupling}\right)^2\log\left[1+\frac{k\left(k-2k_{Coupling}\right)}{k_{Coupling}^2}\left(\frac{\omega}{\omega_C}\right)^2\right]\\ -k\left(k-2k_{Coupling}\right)\left(\frac{\omega}{\omega_C}\right)^2\end{array}\right\} \quad (4)$$

One can easily show that for a fixed $k$ and $k_{Coupling}$, this function is a monotonic increasing function of $\omega$ over the interval $[0,\omega_C]$. On the other hand, for a fixed frequency $\omega$, $\tau_{3D}$ manifests a maximum at $k_{Coupling}=k$. Actually, one can prove in the general case of 3D point junction between two dissimilar solid materials that the maximum transmission function occurs at exactly $k_{Coupling} = 2k_1k_2/(k_1+k_2)$ where $k_1$, $k_2$ denote the spring constants of the two solid materials, respectively.

One should note also that because of the integration over $q$ in the expression of $\tau_{3D}$, the values of the latter might be higher than *1*. In that regard, $\tau_{3D}$ cannot be considered as a transmission coefficient in a proper physical sense. By keeping this in mind, we will, however, continue to address it as such in the rest of this paper.

The combination of Eqs. (1) and (4), allows obtaining the final expression of the phononic thermal conductance $\sigma_{Ph}$ through the point junction between two identical semi-infinite parallel plane solid materials coupled via a certain phonon coupling mechanism.

### III. RESULTS AND DISCUSSION

#### a. *General phonon coupling mechanism*

We start this section by discussing some general features of the transmission function $\tau_{3D}\left(\omega^2,\omega_C^2,k,k_{Coupling}\right)$. The derived expression of the latter function as given by Eq. (4), for the present geometrical configuration of a point junction between two identical isotropic semi-infinite parallel plane solid media, a separation distance $d$ apart, is an exact general expression within the framework of the SBM regardless the nature of spring coupling constant in between.

This expression captures very well the physics of phonon induced transport through the point junction and leads to the correct asymptotic behaviors when $k_{Coupling} \to 0$ and $k_{Coupling} \to +\infty$. In the first case, which corresponds to weak coupling, $\sigma_{Ph}(T,\omega_C,k_{Coupling} \to 0) \to 0$, while in the second case corresponding to strong coupling, $\sigma_{Ph}$ saturates at a value given by:

$$\sigma_{Ph}^{Sat}(T,\omega_C,k_{Coupling} \to +\infty) = \frac{1}{8\pi^2 v^2} \int_0^{\omega_C} \left[\omega^2 - \frac{\omega^4}{2\omega_C^2}\right] C_{Ph}(\omega,T) d\omega \quad (5)$$

Eq. (5) is exactly what one obtains for $\sigma_{Ph}$ using the phonon radiation model.[8] In the low temperature regime, one can replace $\omega_C$ by infinity and the integral can be calculated exactly. This leads to the well-know $T^3$ power law in analogy to the black body photon radiation:[8]

$$\begin{cases} \sigma_{Ph}(T \to 0) = 4 S_{Ph}^B T^3 \\ S_{Ph}^B = \dfrac{\pi^2 K_B^4}{120 \hbar^3 v^2} \end{cases} \quad (6)$$

where $S_{Ph}^B$ represents the phonon Stefan-Boltzmann constant.[8]

At the maximum value of $\tau_{3D}$, obtained when $k_{Coupling}=k$, $\sigma_{Ph}$ reaches its maximum value too. The latter is given by:

$$\sigma_{Ph}^{Max}(T,\omega_C,k_{Coupling}=k) = \frac{1}{8\pi^2 v^2} \int_0^{\omega_C} \omega^2 C_{Ph}(\omega,T) d\omega \quad (7)$$

By comparing Eqs. (5) and (7), one can see that the difference between the maximum and saturation values tends to disappear in the low temperature regime. In this regime, the maximum and saturation values merge to one single value which is obtained not at $k_{Coupling}=k$, but even before at $k_{Coupling}=k/2$. As a matter of fact, one can straightforwardly show that:

$$\sigma_{Ph}(T,\omega_C,k_{Coupling} \to k/2) = \sigma_{Ph}(T,\omega_C,k_{Coupling} \to +\infty) \quad (8)$$

In the high temperature regime, the general expression of the phononic thermal conductance $\sigma_{Ph}$ can be simplified, on using the high temperature expression of $C_{Ph}(\omega,T) \simeq K_B$ where all phonon modes will be in a highly thermally excited state. In this case, after inserting the expression of $\tau_{3D}(\omega^2,\omega_C^2,k,k_{Coupling})$ as given by Eq. (4) into Eq. (1), the integration over $\omega$ in the latter can be performed analytically and we obtain a closed-form expression of $\sigma_{Ph}(T,\omega_C,k_{Coupling})$:

$$\sigma_{Ph}(T,\omega_C,k_{Coupling}) = \frac{K_B \omega_C^3}{8\pi^2 v^2} \frac{\kappa^2}{(2\kappa-1)^2} \left\{ \frac{2\kappa-1}{3} + (\kappa-1)^2 \left[ \frac{2\kappa ArcCoth\left[\frac{\kappa}{\sqrt{2\kappa-1}}\right]}{\sqrt{2\kappa-1}} + \log\left[(\kappa-1)^2\right] - 2(\log\kappa + 1) \right] \right\} \quad (9)$$

where $\kappa = k_{Coupling}/k$.

In the following, we will analyze the case of a specific phonon coupling mechanism between the two identical isotropic semi-infinite parallel plane solid media; coupling through the dispersion force of Casimir in vacuum. In this case we will note $k_{Coupling} = k_{Casimir}$.

### *b. Coupling through Casimir force*

We will compare the phononic thermal conductance due to this coupling mechanism to the Near Field Radiative Heat Transfer (NFRHT) coefficient due to the contribution of the evanescent waves of the p-polarized electromagnetic (EM) field within the framework of a local dielectric permittivity function theory. As a matter of fact, assuming a local dielectric permittivity function where the latter depends only on the frequency of the EM field, previous investigations of the NFRHT for the same geometrical configuration as above, have shown that for separation distances $d$ much smaller than the dominant thermal wavelength $\lambda_T$ and independently of the material nature (metallic or dielectric), the contribution of s and p polarizations of the evanescent EM waves to the NFRHT coefficient manifest, individually, the same behavior with regard to the separation distance $d$ between the two solid materials. As mentioned in the introduction, the contribution of the evanescent waves increases the NFRHT to become orders of magnitude higher than the black body limit. While, the contribution of the s-polarization saturates as $d$ gets shorter, the contribution of the p-polarization, on the other hand, keeps increasing with decreasing $d$ and tends to follow a $d^{-2}$ law for very small $d$-regime.[3,4] We shall note here that, while the contribution of the p-polarization dominates the NFRHT coefficient for dielectrics, due to the presence of magnetic effects, it is the s-polarization contribution that dominates for metals. In our analysis, we will consider only the case of dielectrics. Therefore, one can write for the NFRHT coefficient:

$$h_r^{Evanp}(d,T) \simeq \int_0^\infty h_{Evan}^p(d,T,\omega) d\omega \quad (10)$$

where $h_{Evan}^p$ represents the p-polarized spectral NFRHT coefficient, the expression of which can further be simplified assuming the electrostatic limit to be valid in the small $d$-regime. One can show that, in this case, $h_r^{Evanp}(d,T)$ takes a closed-form expression using the polylogarithm function of second order:[20]

$$\begin{cases} h_r^{Evanp}(d,T) = \dfrac{\delta G(T)}{d^2} \\ \delta G(T) = \dfrac{3}{2\pi^3} g_0 \int_0^\infty h^0(u) \dfrac{\text{Im}^2[r_P(u)]}{\text{Im}[r_P^2(u)]} \text{Im}\{Li_2[r_P^2(u)]\} du \end{cases} \quad (11)$$

In Eq. (11), $g_0 = \pi K_B^2 T / 6\hbar$ is the quantum of thermal conductance, $h^0(u) = u^2 e^u / (e^u - 1)^2$ and $r_P(u) = [\varepsilon(u) - 1]/[\varepsilon(u) + 1]$ represents the Fresnel reflection coefficient of the p-polarized evanescent EM wave in the electrostatic limit.[20]

According to Lifshitz[21] and Schwinger[22] theories of Casimir force, the latter is temperature dependent in general, but as affirmed by many studies, the explicit thermal corrections, even in the high temperature regime, can be neglected when the separation distance $d$ is very small in comparison to the dominant thermal wavelength $\lambda_T$.[23] Since this is the $d$-regime, we are interested to in our study, we will therefore use the zero-temperature expression of Casimir force in the small separation regime. According to Lifshitz, Casimir force per unit area takes a very compact expression in this regime, independently of the nature of the materials under study:[21,24]

$$F_{Casimir}(d, T=0) = \dfrac{\hbar}{16\pi^2 d^3} \int_0^\infty \left\{ \int_0^\infty \dfrac{x^2}{\left[\dfrac{\varepsilon(iy)+1}{\varepsilon(iy)-1}\right]^2 e^x - 1} dx \right\} dy$$

$$= \dfrac{\hbar}{8\pi^2 d^3} \int_0^\infty Li_3[r_P^2(iy)] dy \quad (12)$$

where $Li_3$ is the polylogarithm function of order 3 and $r_p$ is Fresnel reflection coefficient of the p-polarized EM wave in the electrostatic limit as introduced in the expression of $h_r^{Evanp}(d,T)$ in Eq. (11). One should note here that there still is an implicit temperature dependence of Casimir force through $r_p$.

The Casimir spring coupling constant is defined as the absolute value of the derivative of Casimir force per unit area with respect to the separation distance $d$, multiplied by the lattice constant squared. Thus, we get:

$$k_{Casimir}(d) = \left|\frac{\partial F_{Casimir}}{\partial d}\right| a^2 = \frac{3\hbar a^2}{8\pi^2 d^4} \int_0^\infty Li_3\left[r_P^2(iy)\right]dy \quad (13)$$

where $a$ denotes the lattice constant of the solid medium.

**Table 1 :** Physical and geometrical properties of the different materials.

| Material | Lattice constant (Å) | Atomic mass (×10⁻²⁶ kg) | Longitudinal sound velocity (m/s) | Transverse sound velocity (m/s) | Spring constant (N/m) |
|---|---|---|---|---|---|
| **Si** | 5.431[a,b] | 4.66[a,b] | 8430[b] | 5640[b] | 6.16[d] |
| **3C-SiC** | 4.36[b] | 3.33[c] | 9500[b] | 4100[b] | 4.04[d] |

a: Reference [25].

b: Reference [26].

c: Calculated as $m_{SiC} = (m_{Si} + m_C)/2$ where the mass of a single atom of Carbone is $m_C \approx 2\times 10^{-26} kg$.

d: Calculated using the long wavelength approximation for the 1D atomic harmonic chain dynamics as: $k = mv^2/a^2$.[25]

In order to illustrate our results, we consider two dielectrics (Si and SiC) as typical materials. In addition, SiC is taken to be in a cubic crystallographic configuration (3C-SiC). We will consider the temperature to range from *300K* to *800K*. The needed physical and geometrical properties of the two materials are given in Table 1.

Si will be assumed to be highly n-doped with a doping level ranging from *10¹⁸cm⁻³* to *10²¹cm⁻³*. The dielectric permittivity function of Si is described using Drude model while the one of SiC is modeled using Lorentz-Drude Model:[3,4]

$$\begin{cases} \varepsilon_{Drude}(\omega) = \varepsilon_b - \dfrac{\omega_p^2}{\omega^2 + i\omega\gamma} & (a) \\ \varepsilon_{Lorentz-Drude}(\omega) = \varepsilon_b \left\{\dfrac{\omega_{LO}^2 - \omega^2 + i\gamma\omega}{\omega_{TO}^2 - \omega^2 + i\gamma\omega}\right\} & (b) \end{cases} \quad (14)$$

where $\varepsilon_b$ is the high frequency dielectric constant that accounts for the bound electron contribution from the bulk, $\omega_p$ is the plasma frequency, $\omega_{LO}$ and $\omega_{TO}$ are the longitudinal and transverse optical phonon frequencies, respectively and $\gamma$ denotes the damping factor. For Si, $\omega_p$ and $\gamma$ are functions of temperature and doping concentration. They are, respectively, given by:[27]

$$\begin{cases} \omega_p^2(N) = \dfrac{Ne^2}{m^*\varepsilon_0} \\ \gamma(N,T) = \dfrac{e}{m^*\mu_e(N,T)} \\ \mu_e(N,T) = 88T_n^{-0.57} + \dfrac{7.4\times10^8 T^{-2.33}}{1+(0.88/1.26)10^{-17}NT_n^{-2.546}} \\ T_n = T/300 \end{cases} \quad (15)$$

where $e$ is the electron elementary charge, $m^*=0.27m_0$ is the electron effective mass, $m_0$ is the electron rest mass, $N$ denotes the doping concentration, $\mu_e$ is the electron mobility and $\varepsilon_0$ represents the vacuum permittivity. We note here that $m^*$ is assumed to be temperature independent.

Over the temperature interval considered *[300-800K]*, $\omega_{LO}$ and $\omega_{TO}$ of SiC change by less than *2%* and hence, they are taken to be constants; $\omega_{LO}=1.826\times10^{14}\,\text{rad.s}^{-1}$ and $\omega_{TO}=1.495\times10^{14}\,\text{rad.s}^{-1}$. On the other hand, the damping factor $\gamma$ increases linearly with temperature $\gamma(T)=1.885\times10^{11}\left[4.8329+0.0183(T-300)\right]\,\text{rad.s}^{-1}$.[20]

Because of the smallness of the thermal expansion coefficient ($\sim10^{-5}\text{K}^{-1}$) for both materials, we can neglect the temperature dependence of the intrinsic spring coupling constant $k$. In addition, we can easily check that the equivalent Debye-like temperatures ($\theta_C^D = \hbar\omega_C/K_B$) corresponding to the phonon cutoff frequencies of the two materials (~176K for Si) and (~168K for 3C-SiC) are almost half the room temperature (300K). Hence, we have all the conditions to use the closed-form high temperature expression of the phononic thermal conductance $\sigma_{Ph}$ as given by Eq. (9), in which $k_{Coupling}$ is replaced by $k_{Casimir}$ and $\kappa \equiv k_{Casimir}(d)/k$.

In Fig. 2, we report the variation of Casimir spring coupling constant $k_{Casimir}$ for both highly n-doped Si and 3C-SiC as a function of the gap distance $d$ at room temperature. For the numerical calculation of $k_{Casimir}$, the integration in the angular frequency domain is taken to vary from zero to $10K_BT/\hbar$ in a similar way as for the numerical calculation of the NFRHT coefficient $h_r^{Evanp}(d,T)$, since this integration is governed by the Planck spectrum emission band. As a matter of fact, the amplitude of the Planck spectrum falls down to less than 5% of its maximal value at $\omega=10K_BT/\hbar$.

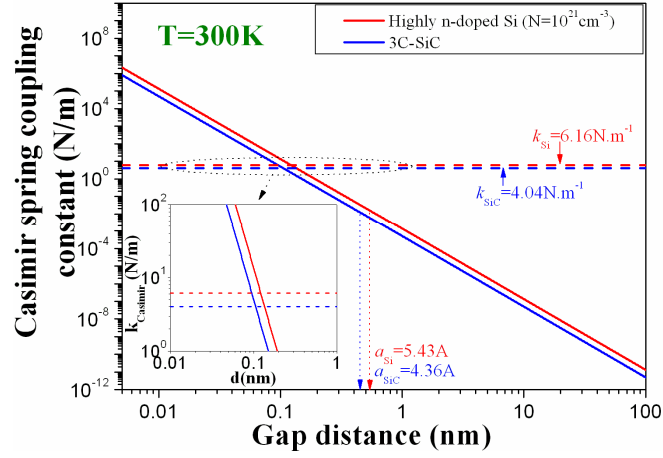

**Figure 2 :** Room temperature behavior of the Casimir spring coupling constant for both highly n-doped Si ($N=10^{21}cm^{-3}$) and 3C-SiC as a function of the gap distance.

Numerical simulations have shown the effect of temperature and doping concentration to be negligible over the studied interval *[300-800K]*. We pushed down the gap distance *d* to the picometer range, which rigorously speaking, has no physical sense. The reason is nevertheless, threefold; (i) to show the huge sensitivity of $k_{Casimir}$ to *d*, (ii) to point out the value of *d* at which $k_{Casimir}=k$ and (iii) to show that the phononic thermal conductance $\sigma_{Ph}(d,T)$ saturates mathematically as $d \to 0$. As one can see in Fig. 2, Casimir spring coupling constants of both materials take very close values as functions of *d*. Moreover, for both materials, $k_{Casimir}=k$ lies at distances *d~1Å*, much smaller than the lattice constant.

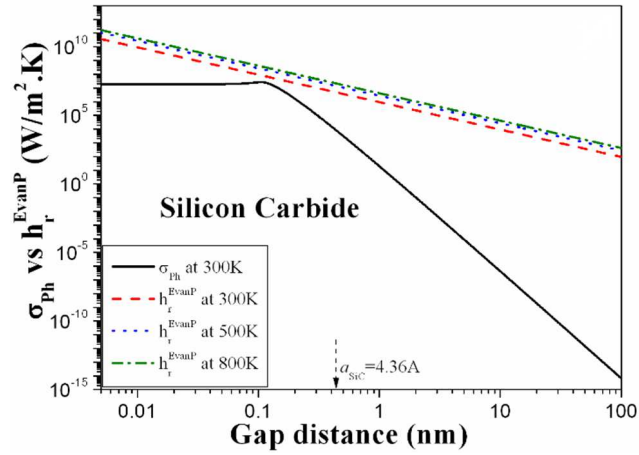

**Figure 3 :** Behavior of the phononic thermal conductance and the NFRHT coefficient as functions of the gap distance through a point junction between two identical isotropic semi-infinite parallel plane solid media of 3C-SiC.

Figures 3 and 4(a-d) illustrate a comparison between the calculated $\sigma_{Ph}(d,T)$ and the NFRHT coefficient $h_r^{Evanp}(d,T)$ through a point junction between two identical isotropic semi-infinite parallel plane solid media of 3C-SiC and highly n-doped Si, respectively. For both materials, $\sigma_{Ph}(d,T)$ turns out to be less sensitive to temperature $T$ and doping concentration $N$ for the values considered above of highly n-doped Si. On the other hand, $h_r^{Evanp}(d,T)$ appears to be sensitive to both $T$ and $N$ for Si and to $T$ for SiC. In addition, in the case of highly n-doped Si, the sensitivity of $h_r^{Evanp}(d,T)$ to $T$ proves to be dependent on $N$. Thus only the room temperature $\sigma_{Ph}(d,T=300K)$ is represented for both dielectrics.

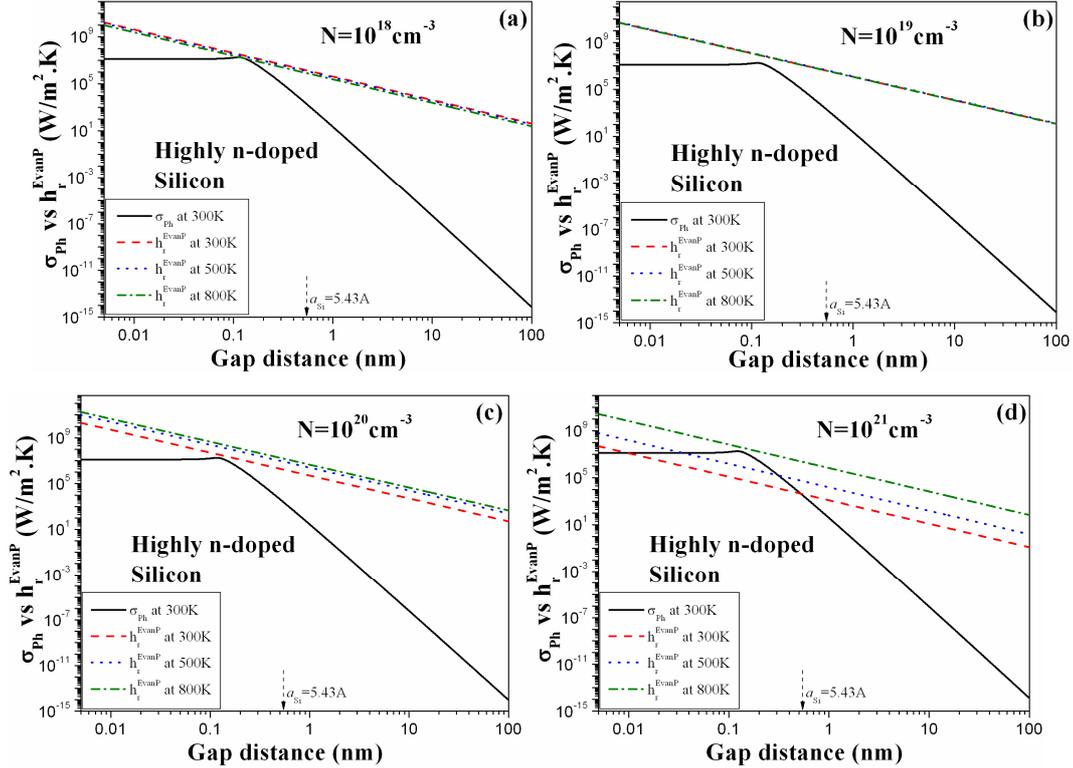

**Figure 4 :** Behavior of the phononic thermal conductance and the NFRHT coefficient as functions of the gap distance through a point junction between two identical isotropic semi-infinite parallel plane solid media of highly n-doped Si at different doping levels.

Starting from Eq. (9), one can straightforwardly check that $\sigma_{Ph}(d,T=300K)$ reaches a maximum value of $\sigma_{Ph}^{Max} = K_B \omega_C^3 / 24\pi^2 v^2$ at $\kappa = k_{Casimir}/k = 1$ and tends to a saturation value of $\sigma_{Ph}^{Sat} = 7K_B \omega_C^3 / 240\pi^2 v^2$ when $\kappa = k_{Casimir}/k \to +\infty$ which corresponds to $d \to 0$. Therefore the ratio between the maximum and the saturation values of $\sigma_{Ph}(d,T=300K)$ is exactly $R = \sigma_{Ph}^{Max}/\sigma_{Ph}^{Sat} = 10/7$.

From the above figures, we see that the NFRHT dominates the heat transfer in the case of 3C-SiC, except around the distance at which $\sigma_{Ph}(d,T=300K)$ reaches a maximum where Induced Phonon Transfer through Casimir Force (IPTTCF) becomes a non-negligible fraction of the total heat transfer process. In the case of highly n-doped Si, we first notice that for a fixed temperature $T$, the NFRHT coefficient manifests a maximal value at an optimal doping level $N_{opt}$. This behavior for highly n-doped Si has already been shown previously.[28] The interplay between IPTTCF and NFRHT depends primarily on the doping level $N$ then secondarily on $T$. IPTTCF starts to dominate the heat transfer as the doping level increases. One can see that for $N=10^{21}cm^{-3}$, the transition distance lies around the lattice constant at room temperature and tends to decrease by increasing the ambient temperature. As one increases the value of $N$, the dielectric material electrical behavior approaches the one of a semi-metallic material. It is known, on the other hand that metals manifest the highest Casimir force values.[14,15] and the lowest NFRHT coefficients values.[3,4]

It is worth noting to mention here the difference between the approach of Budaev and Boggy[12,13] and the approach presented herein. The former authors found a transition distance for Si at room temperature of ~$5nm$, almost $10$ times higher than the one we found above. This value appears to be a large separation distance though. We believe the reason of this huge disagreement lies mainly in the very heuristic treatment followed by the authors, particularly the use of grossly approximated formula of different quantities involved in the estimation of the heat transfer fluxes, which have led to an overestimation of the effect of IPTTCF.

The transition distance below which IPTTCF dominates NFRHT is of the order or smaller than the lattice constant. This, makes the domain of validity of our herein presented approach undoubtedly narrow. It however and certainly shows that IPTTCF constitutes a plausible and a very potential mechanism to capture and describe the natural transition from the radiative regime to the conductive regime of heat transfer. The IPTTCF mechanism would even be enhanced if combined to other potential coupling mechanisms such as charge-charge electrostatic interaction or piezoelectricity that was recently analyzed by Prunnila and Meltaus.[11] These different mechanisms could simply be included by attributing adequate expressions for the spring coupling constant $k_{Coupling}$.

For gap distances of the order or less than the lattice constant $(d \leq a)$, the microscopic variation and the discrete character of the matter will take over the continuum approximation. Thus, one expects other additional effects to come into play and even to be more dominant, mainly nonlocal effects of the dielectric permittivity function[29,30] as well as quantum electronic coupling effects[31] especially at distances of the order or smaller than the interatomic distances which for Si is about $0.24nm$ at room temperature.

## VI.  SUMMARY

Induced Phonon Transfer in vacuum constitutes a potential mechanism to describe the natural transition from the radiative regime to the conductive regime of heat transfer at the point junction interface between two identical solid dielectric materials when the distance gap between the latter becomes very small so that they can mutually touch. We specifically studied how this induced phonon transfer could be mediated by Casimir force. We showed that this transfer could become dominant when the distance gap becomes of the order or smaller than the lattice constant of the dielectric material. However, at these distances, one expects other additional effects to come into play and even to be more dominant, mainly nonlocal effects of the dielectric permittivity function as well as quantum electronic coupling effects, particularly at distances of the order or smaller than the interatomic distances. Hence, a full and complete study of the natural transition from the radiative regime to the conductive regime of heat transport will certainly necessitate taking into account all these effects in a more elaborate and sophisticated theory that includes all possible coupling mechanisms depending on the materials nature (metallic or dielectric, similar or dissimilar) and their surface state and which will go beyond the fluctuational electrodynamics theory based on which, the derived expressions for the Casimir force and the NFRHT coefficient were used in the present study. This study will also bring to light key information about the fundamental behavior of the solid-solid interface thermal resistance.